\documentclass[12pt]{iopart}

\usepackage[utf8]{inputenc}
\usepackage{amsbsy}
\usepackage{amssymb}
\usepackage{graphicx}
\usepackage[colorlinks=true,citecolor=blue]{hyperref}
\usepackage{units}
\usepackage{color}

\newcommand{\text}[1]{\mathsf{#1}}
\newcommand{\kBT}{k_{\rm B}T}
\newcommand{\kB}{k_{\rm B}}
\newcommand{\bea}{\begin{eqnarray}}
\newcommand{\eea}{\end{eqnarray}}
\newcommand{\be}{\begin{equation}}
\newcommand{\ee}{\end{equation}}

\begin{document}

\title[Heat diode and engine based on quantum Hall edge states]{Heat diode and engine based on quantum Hall edge states}

\author{Rafael S\'anchez}
\address{Instituto de Ciencia de Materiales de Madrid (ICMM-CSIC), Cantoblanco 28049, Madrid, Spain}
%\ead{rafael.sanchez@icmm.csic.es}
\author{Bj\"orn Sothmann}
\address{D\'epartement de Physique Th\'eorique, Universit\'e de Gen\`eve, CH-1211 Gen\`eve 4, Switzerland}
\author{Andrew N. Jordan}
\address{Department of Physics and Astronomy, University of Rochester, Rochester, New York 14627, U.S.A.}
\address{Institute for Quantum Studies, Chapman University, Orange, California 92866, U.S.A.}
\begin{abstract}
We investigate charge and energy transport in a three-terminal quantum Hall conductor. 
% The peculiar properties introduced by a strong magnetic field in the form of chiral propagation along the edges of the sample have important consequences in the response to thermal biases.
The peculiar properties of chiral propagation along the edges of the sample have important consequences on the response to thermal biases.
Based on the separation of charge and heat flows, thermoelectric conversion and heat rectification can be manipulated by tuning the scattering at gate-modulated constrictions. Chiral motion in a magnetic field allows for a different behaviour of left- and right-moving carriers giving rise to thermal rectification by redirecting the heat flows. We propose our system both as an efficient heat-to-work converter and as a heat diode.
\end{abstract}

\maketitle

\section{Introduction}
The quantum Hall effect is a paradigm of mesoscopic transport since its discovery in 1980~\cite{v._klitzing_new_1980}. In the presence of a strong perpendicular magnetic field, electrons are restricted to move along the edges of a two-dimensional electron gas~\cite{halperin_quantized_1982}. Multiterminal experiments reveal quantized plateaus in the Hall resistance. These can be explained within Landauer-B\"uttiker theory by the motion of electrons along edge states not affected by back-scattering~\cite{buttiker_absence_1988}. A huge interest has been devoted to the transport of charge for potential metrological applications as well as  for the possibility to use the chirality of edge channels to construct quantum optics interferometers for electrons~\cite{ji_electronic_2003}.

Much less attention has been paid to the transport of energy in such systems. This comes in spite of its fundamental relevance for examining the role of energy relaxation~\cite{granger_observation_2009,nam_thermoelectric_2013} and interactions~\cite{altimiras_non-equilibrium_2010} within quantum Hall conductors. The experiments by Granger {\it et al.}~\cite{granger_observation_2009} and Nam {\it et al.}~\cite{nam_thermoelectric_2013} introduced the thermoelectric response in multiterminal setups as an efficient way to probe the chirality of motion and the thermalization of carriers. Boosted by the recent interest in thermodynamics in the quantum regime, the thermoelectric properties of quantum Hall systems have been investigated, including the thermopower of a quantum point contact~\cite{dambrumenil_thermopower_2013} and a Mach-Zehnder interferometer~\cite{hofer_quantum_2015}, the Nernst effect~\cite{checkelsky_thermopower_2009,sothmann_quantum_2014}, nonlinear effects~\cite{hwang_magnetic-field_2013,lopez_thermoelectric_2014}, and spin-Hall systems~\cite{rothe_spin-dependent_2012,dolcetto_generating_2013,hwang_nonlinear_2014}. Remarkably it has allowed for the measurement of the quantum of heat conductance~\cite{pendry_quantum_1983,jezouin_quantum_2013}. Most of these works discuss two-terminal configurations. 

\begin{figure}[t]
\begin{center}
\includegraphics[width=0.65\linewidth,clip]{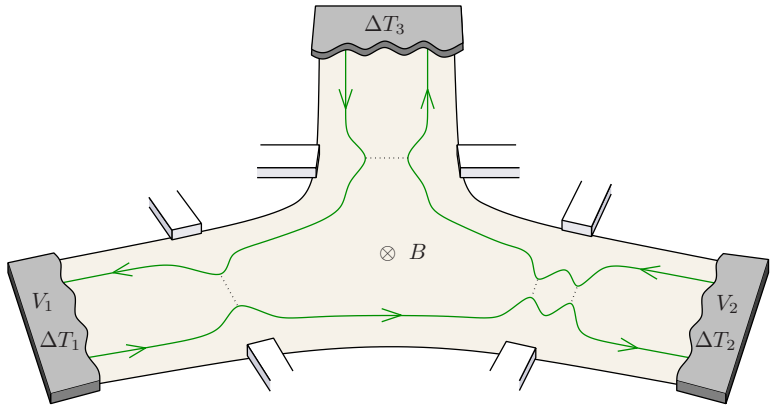}
\end{center}
\caption{\label{scheme} Three-terminal quantum Hall thermoelectric device. Current flows between terminals 1 and 2 when a temperature gradient is applied either longitudinally (at terminals 1 or 2) or transversally (at terminal 3). A voltage $V=V_1-V_2$ applied against the generated current allows to extract useful power. Terminal 3 is considered as a probe which does not inject charge into the system.  The thermoelectric response relies on the energy dependence of the scattering at the constrictions, in this case a quantum point contact in terminal 1, and a resonance in terminal 2.}
\end{figure}

The importance of the quantum Hall regime appears however in multiterminal samples. In particular, we will focus on the minimal model containing three terminals: two of them support a charge current with the third one acting as a voltage probe, see \fref{scheme}. In contrast to two-terminal geometries, three-terminal devices enable a separation of charge and energy flows: one of the contacts is used to inject heat but no charge into the system. The principle of the separation of currents allows one to manipulate them individually. Thus different functionalities can be defined for our three-terminal device. 

On the one hand, the conversion of the injected heat flows can be used to generate a finite electrical power. Then the system works as a heat engine.  
Multi-terminal heat engines have been proposed with the coupling to the hot bath being mediated by a capacitance~\cite{sanchez_optimal_2011,sothmann_rectification_2012}, electron-boson interaction~\cite{entin-wohlman_three-terminal_2010,entin-wohlman_three-terminal_2012,sothmann_magnon-driven_2012,ruokola_theory_2012,jiang_three-terminal_2013,bergenfeldt_hybrid_2014,entin-wohlman_enhanced_2015}, or a central cavity~\cite{jordan_powerful_2013,sothmann_powerful_2013,whitney_most_2014}. A finite charge current is then generated between the other two contacts by thermoelectric energy conversion. Quite generally, the converter must be such that left-right and particle-hole symmetries are broken. This establishes a preferred direction for a net current to flow. Recent experiments have demonstrated this mechanism~\cite{roche_harvesting_2015,hartmann_voltage_2015}.

On the other hand, the redistribution of heat flows among the three terminals leads to thermal rectification. Applying opposite temperature gradients leads to different thermal responses. When the ratio of the two currents is large enough, the system behaves as a heat diode, a key element for the control of heat in electronic devices~\cite{li_colloquium:_2012}. This effect has been shown to be prominent in mesoscopic conductors~\cite{li_thermal_2004,segal_spin-boson_2005,chang_solid-state_2006,scheibner_quantum_2008,ruokola_single-electron_2011,martinez-perez_rectification_2014}. We show that the physics of the quantum Hall effect can be used to construct a perfect heat diode in the linear regime.

In a recent work, we investigated the unique thermoelectric properties of a three-terminal configuration in the quantum Hall regime~\cite{sanchez_chiral_2014}.  There, a novel contribution to the three-terminal thermopower appears due to the chiral propagation along edge states. This term is responsible for a finite thermoelectric response in left-right symmetric configurations, which is not present in time-reversal symmetric systems. Furthermore, the presence of a thermoelectric response depends on which contact the electron-hole symmetry is broken at. For example in the configuration depicted in \fref{scheme} with heat injected from terminal 3, current will flow only if scattering at the left junction is energy dependent. For a reversed magnetic field, the same is true for the junction on the right. As a consequence of the Onsager reciprocity relations~\cite{onsager_reciprocal_1931}, this property gives rise to an extreme asymmetry between the Seebeck and Peltier coefficients. 

Here, we will further explore the properties of chiral charge and heat transport in this configuration. We consider that electrons keep their phase-coherence in their transport between the different terminals. This is the case in quantum Hall samples at low temperatures (${\sim}\unit[100]{mK}$), where the contribution of electron-phonon coupling~\cite{zeitler_investigation_1993,tieke_fundamental_1997} is negligible.  We will distinguish the different behaviour of the {\it longitudinal} and {\it crossed} responses introduced by the three-terminal geometry. The former ones are due to a temperature bias applied in one of the terminals that carry a charge current. They correspond to a two-terminal conductor with a probe coupled to it~\cite{sanchez_thermoelectric_2011}. We will show that chirality produces a huge heat rectification effect in the longitudinal terms in the presence of a probe. Hence, our system works as an efficient heat diode. The crossed term is due to the heating of the probe terminal. This is the mechanism of interest for energy harvesting~\cite{sothmann_thermoelectric_2015}. After introducing the scattering formalism in section~\ref{scattering}, the properties of different configurations will be analyzed in sections~\ref{sec:qpc} and ~\ref{sec:restunn}, including the thermoelectric 
 performance of the corresponding configurations working as a heat diode and a heat engine. Conclusions are presented in section~\ref{sec:conclusions}.

%%%%%%%%%%%%%%%%%%%%%
\section{Scattering theory}
\label{scattering}
%%%%%%%%%%%%%%%%%%%%%

Assuming non-interacting electrons, electronic transport through the system is well described by the Landauer-B\"uttiker formalism~\cite{buttiker_absence_1988}. We will restrict ourselves to the linear-response regime where the electrical and thermal gradients are small compared to other energy scales in the system. Extensions to the non-linear regime have been addressed recently~\cite{hwang_magnetic-field_2013}. Charge and heat currents ${\mathbf I}_i=(I_i^e,I_i^h)$ are given by the transmission probabilities ${\cal T}_{i\leftarrow j}(E)$ for electrons injected in terminal $j$ to be absorbed by terminal $i$ and the electric and thermal affinities ${\mathbf F}_j=(F_j^V,F_j^T)$,
\be
\label{Ii}
{\mathbf I}_i=
\frac{1}{h}\sum_j\int dE[N_j\delta_{ij}-{\cal T}_{i\leftarrow j}(E)]\xi(E)
\left(
	\begin{array}{cc}
		e & eE \\
		E & E^2
	\end{array}
\right)
{\mathbf F}_j,
\ee
with $F_i^V=eV_i/(\kBT)$ and $F_i^T=k_\text{B}\Delta T_i/(\kBT)^2$,
where $V_i$ and $\Delta T_i$ are the voltage and temperature bias applied to terminal $i=1,2,3$, respectively, and $\kBT$ is the system temperature~\cite{buttiker_four-terminal_1986,butcher_thermal_1990,mazza_thermoelectric_2014}. $N_i$ is the number of channels in terminal $i$. In our case, we will consider $N_i=1$ for all leads. We have defined $\xi(E)=-(\kBT/2)df(E)/dE$, with the Fermi function $f(E)$. We choose the equilibrium Fermi energy $E_{\rm F}$ as the zero of energy.

We are mainly interested in the configuration where terminal 3 injects heat but no charge, i.e. it acts as a voltage probe~\cite{buttiker_coherent_1988,bergfield_thermoelectric_2014}. Hence, we solve Eqs.~\eref{Ii} to get the voltage developed at the probe satisfying the boundary condition $I_3^e=0$. From charge conservation we thus have $I^e=I_1^e=-I_2^e$. Note that by gauge invariance the response to the electrical affinities depends only on the voltage difference applied to the conducting terminals, $V_1-V_2$. We can define the Onsager coefficients ${\mathbf {\cal L}}$ with 
\be
\left(
	\begin{array}{c}
		I^e \\
		I_j^h
	\end{array}
\right)
=
\left(
	\begin{array}{cc}
		{\cal L}_{11}^{eV} & {\cal L}_{1i}^{eT} \\
		{\cal L}_{j1}^{hV} & {\cal L}_{ji}^{hT}
	\end{array}
\right)
\left(
	\begin{array}{c}
		F_1^V{-}F_2^V \\
		F_i^T
	\end{array}
\right).
\ee
The diagonal term ${\cal L}_{11}^{eV}$ corresponds to the charge conductance while the diagonal terms ${\cal L}_{ji}^{hT}$ are related to the various heat conductances that occur in the setup. The thermoelectric response is contained in the off-diagonal terms in the first row, ${\cal L}_{1i}^{eT}$, and in the first column, ${\cal L}_{j1}^{hV}$. These are related to the Seebeck (the electric response to a thermal gradient) and the Peltier (the thermal response to an electric field) coefficients, respectively.
Due to the presence of a magnetic field that breaks time-reversal symmetry, the Onsager reciprocity relations~\cite{onsager_reciprocal_1931,buttiker_symmetry_1988,butcher_thermal_1990,matthews_experimental_2014} for our setup read
% In the absence of a magnetic field, Onsager reciprocity relations imply ${\cal L}_{j1}^{eT}=e{\cal L}_{1j}^{hV}$. Importantly for us, this relation holds in the presence of a magnetic field under the reversal of the field~\cite{onsager_reciprocal_1931,buttiker_symmetry_1988,butcher_thermal_1990,matthews_experimental_2014}:
\be
{\cal L}_{j1}^{eT}(B)=e{\cal L}_{1j}^{hV}(-B).
\ee
Note that in the linear regime not only charge currents but also heat currents must be conserved, i.e. $\sum_jI_j^h=0$.

By introducing scatterers such as quantum point contacts or resonances into the leads, cf. \fref{scheme}, we obtain nontrivial transmission probabilities ${\cal T}_{i\leftarrow j}(E)$. In general, they will depend on energy which is a key requirement to obtain a finite thermoelectric response.
The effects that we want to discuss in the following are most dramatic when scattering does not occur in all terminals. Otherwise, a closed loop forms in the center of the sample which removes the effect of chiral propagation. To be more specific, we will focus on the case when the coupling of the probe terminal to the system is transparent. A comparison to configurations with closed orbits is given in \ref{sec:closed} and \ref{sec:antidot}.
 
We choose a positive magnetic field penetrating the sample as in \fref{scheme}. Then, the propagation between the different terminals is given by the transmission probabilities ${\cal T}_i(E)$ at the different junctions: ${\cal T}_{1\leftarrow3}(B)={\cal T}_1(E)$, ${\cal T}_{3\leftarrow2}(B)={\cal T}_2(E)$. Note that while ${\cal T}_{2\leftarrow1}(B)={\cal T}_1(E){\cal T}_2(E)$, electrons in the opposite direction are absorved by the probe and ${\cal T}_{1\leftarrow2}(B)=0$. This different behaviour of left and right movers is a direct consequence of the chiral propagation along edge states. When one reverses the magnetic field electrons flow in the opposite direction, so that we have ${\cal T}_{2\leftarrow1}(-B)=0$, ${\cal T}_{3\leftarrow1}(-B)={\cal T}_1(E)$, ${\cal T}_{2\leftarrow3}(-B)={\cal T}_2(E)$.

The electrical response of each junction $l=1,2$ can be separately parametrised by its two-terminal charge conductance, 
\begin{eqnarray}
G_l=\frac{e^2}{\kBT h}\int dE{\cal T}_l(E)\xi(E),
%K_l&=\frac{e^2}{\kBT h}\int dEE^2{\cal T}_l(E)\xi(E),
\end{eqnarray}
and its thermopower, 
\be
S_l=\frac{e}{hk_\text{B}T^2G_l}\int dEE{\cal T}_l(E)\xi(E).
\ee
The latter is defined as the voltage developed across the junction for a given temperature gradient.
Note that the thermopower is finite only in the presence of energy-dependent scattering.
In the case where this dependence is weak, $G_l$ and $S_l$ are related by the Cutler-Mott formula~\cite{cutler_observation_1969,lunde_mott_2005}. Analogously, the junction thermal conductance is
\begin{eqnarray}
%G_l=\frac{e^2}{\kBT h}\int dE{\cal T}_l(E)\xi(E)
N_l=\frac{1}{h(\kBT)^2}\int dEE^2{\cal T}_l(E)\xi(E).
\end{eqnarray}
For later convenience, we define the integrals
\be
g_{l}^{(n)}=\int dE E^{n-1}{\cal T}_l(E)\xi(E),
\ee
\be
j^{(n)}=\int dE E^{n-1}{\cal T}_1(E){\cal T}_2(E)\xi(E),
\ee
the latter describing the elastic propagation between the two conductor terminals without losing the phase coherence. %Finally, $M^{(n)}=g_{1}^{(n)}+g_{2}^{(n)}-j^{(n)}=(g_{1}^{(n)}+g_{2}^{(n)})\lambda^{(n)}$. 
Its influence on the transport coefficients will be parametrised by the factors $\lambda^{(n)}=1-j^{(n)}/(g_{1}^{(n)}+g_{2}^{(n)})$.
 
The charge response to a voltage applied between the conductor terminals simply reads ${\cal L}_{11}^{eV}=k_\text{B}TG/e$, with a conductance 
\be
G=\frac{1}{\lambda^{(1)}}\left(\frac{1}{G_1}+\frac{1}{G_2}\right)^{-1}.
\ee
It differs from the one expected for the sequential transmission through a series of two barriers in a two terminal measurement~\cite{buttiker_coherent_1988} by the factor $\lambda^{(1)}$.% indicating coherent transport along the edge.

The thermal conductance of the hot probe is given by
\be
\label{L33ht}
{\cal L}_{33}^{hT}
%=M_3-\frac{\kB T^2\Lambda_2}{e\Lambda}\frac{G_1S_1+G_2S_2}{G_1+G_2}J_2
%=M_3-\frac{\kB T^2}{e}\frac{G}{G_1G_2}\Lambda_2(G_1S_1+G_2S_2)J_2,
%=(\kBT)^2(N_1+N_2)-\frac{\kB T^2}{e}\frac{G}{G_1G_2}\Lambda_2(G_1S_1+G_2S_2)J_2-\frac{1}{e^2\kBT}J_3,
=(\kBT)^2(N_1+N_2)\lambda^{(3)}-\kB T^3\frac{[(G_1S_1+G_2S_2)\lambda^{(2)}]^2}{(G_1+G_2)\lambda^{(1)}}.
\ee
%with %$M_3=h^{-1}\int dE E^2[{\cal T}_1(E)+{\cal T}_2(E)]\xi(E)-(e^2\kBT)^{-1}J_3$ and 
%$\Lambda_2=1-(e\kB^2T^3)^{-1}J_2/(G_1S_1+G_2S_2)$. 
Obviously, the flow of heat  depends on the details of the scattering. For instance, if the two junctions are energy independent, the second term in Eq.~\eref{L33ht} vanishes and we get ${\cal L}_{33}^{hT}=L_0T ^2$, where $L_0=(\pi\kB/e)^2/3$ is the Lorentz number. 
%$g_{n,l}=e^2h^{-1}\kBT\int dE E^{n-1}{\cal T}_l(E)\xi(E)$,

\subsection{Thermoelectric response}
For the configuration shown in \fref{scheme}, the thermoelectric coefficients read
\begin{eqnarray}
\label{L11et}
{\cal L}_{11}^{eT}(B)&=k_\text{B}T^2GS_1-e{\cal X}_1,\\
\label{L12et}
{\cal L}_{12}^{eT}(B)&=-k_\text{B}T^2GS_2,\\
\label{L13et}
{\cal L}_{13}^{eT}(B)&=-({\cal L}_{11}^{eT}+{\cal L}_{12}^{eT})=k_\text{B}T^2G(S_2-S_1)+e{\cal X}_1,
\end{eqnarray}
for the Seebeck terms, and
\begin{eqnarray}
\label{L11hv}
{\cal L}_{11}^{hV}(B)&=e^{-1}k_\text{B}T^2GS_1,\\
\label{L21hv}
{\cal L}_{21}^{hV}(B)&=e^{-1}k_\text{B}T^2GS_2+{\cal X}_2,\\
\label{L31hv}
{\cal L}_{31}^{hV}(B)&=e^{-1}{\cal L}_{13}^{eT}-{\cal X}_1-{\cal X}_2,
\end{eqnarray}
for the Peltier terms.
Note that the longitudinal coefficients include a term that is proportional to the thermopower of the junction at which the temperature difference is applied. The crossed terms ${\cal L}_{13}^{eT}$ and ${\cal L}_{31}^{hV}$ depend on the difference of the thermopowers $S_i$. The factor $S_2-S_1$ implies the condition that time-reversal symmetric heat rectifiers must be both left-right and electron-hole asymmetric~\cite{sothmann_thermoelectric_2015}. 

Importantly, for the quantum Hall setup we identify an additional term that introduces deviations from this behaviour. It is due to the chirality introduced by the magnetic field. For the longitudinal terms, it contributes for only one direction of the magnetic field (i.e. it appears in ${\cal L}_{11}^{eT}(B)$ and ${\cal L}_{12}^{eT}(-B)$). However it is unavoidable in the crossed terms. In our quantum Hall conductor, it reads
\be
{\cal X}_l=\frac{1}{h}\frac{GG_{l}}{G_1G_2}(eTS_lj^{(1)}-j^{(2)}).
\label{X}
\ee
For the chosen direction of the magnetic field, ${\cal L}_{13}^{eT}$ depends on ${\cal X}_1$ because it is at this terminal where the electron-hole excitations created in terminal 3 first arrive. For a reversed magnetic field, ${\cal L}_{13}^{eT}(-B)=k_\text{B}T^2G(S_2-S_1)-e{\cal X}_2$ is obtained by Eq.~\eref{L31hv}, as follows from the Onsager relations. Note that the sum of the chiral terms gives a measure of the anti-symmetrized Seebeck response: ${\cal L}_{13}^{eT}(B)-{\cal L}_{13}^{eT}(-B)=e({\cal X}_1+{\cal X}_2)$.

The first important consequence of the chiral term is that a left-right symmetric conductor with $S_1=S_2$ can rectify heat. The reason is that chiral motion defines an asymmetry between the left and right terminals depending on the sign of the applied magnetic field. Note however, that energy-dependent scattering is needed at junction $l$ in order to make ${\cal X}_l$ finite.

A strong signature of chirality is contained in the fact that, if one of the transmissions is constant, one can always chose a direction of the magnetic field such that ${\cal L}_{31}^{hV}=0$. In the case shown in \fref{scheme}, corresponding to Eq.~\eref{L13et}, no heat current will flow in response to an applied bias voltage if ${\cal T}_{1}(E)=c$. This is explained because the non-equilibrium excitations generated around the Fermi level by the hot probe thermalize in terminal 1 if the particle-hole symmetry is not broken at the junction. The same applies to terminal 2 for a reversed magnetic field.

%one either sign of the magnetic field if the transmission at one of the junctions is energy independent. ${\cal L}_{31}^{hV}(B)=0$ if ${\cal T}_1(E)=c$, and ${\cal L}_{31}^{hV}(-B)=0$ if ${\cal T}_2(E)=c$.

As is evident in Eq.~\eref{L13et}, the chiral terms control the asymmetry between the Seebeck and Peltier coefficients. These can be tuned by gating the two junctions. If we parametrize the asymmetries by $x_{ij}={\cal L}_{ij}^{eT}/(e{\cal L}_{ji}^{hV})$, we get, for the longitudinal terms
\bea
x_{11}&=x_{21}=1-\frac{e{\cal X}_1}{\kB T^2GS_1},\\
x_{22}&=x_{12}=\left(1-\frac{e{\cal X}_2}{\kB T^2GS_2}\right)^{-1}.
\eea
% Deviations from $x_{ij}=1$ due to a magnetic field have been proposed to increase the finite-time performance of two-terminal heat engines coupled to a probe~\cite{benenti_thermodynamic_2011,brandner_multi-terminal_2013}. %In particular, there is the prediction that the efficiency at maximum power can exceed the bound of $\eta_c/2$ set for time-reversal symmetric setups, where $\eta_c$ is the Carnot efficiency.
For the crossed terms, we have
\be
x_{13}=1+\frac{e({\cal X}_1+{\cal X}_2)}{k_\text{B}T^2G(S_2-S_1)-e{\cal X}_2}.
\ee
For ${\cal X}_1={\cal X}_2=0$, we recover the limit $x_{ij}=1$ with no magnetic field. Interestingly, our system makes it possible to find configurations where the asymmetry is tuned to be either zero or infinite. 

The thermoelectric performance is usually described in terms of the maximal power 
\be
P_{{\rm m},l}=\frac{\kBT}{4e}\frac{({\cal L}_{1l}^{eT}F_l^T)^2}{{\cal L}_{11}^{eV}},
\ee
extracted at a voltage $V_{{\rm m},l}$ 
when the temperature bias is applied to terminal $l$, and the corresponding efficiency at maximum power:
\be
\eta_{{\rm maxP},l}=\frac{P_{{\rm m},l}}{I_l^h(V_{{\rm m},l})}.
%=\frac{{\cal L}_{1l}^{eT}}{4{\cal L}_{11}^{eV}}x_{1l}\eta_c. 
\ee
For time-reversal symmetric systems, $\eta_{\rm{maxP},l}$ is bounded by $\eta_c/2$ with $\eta_c$ being the Carnot efficiency~\cite{van_den_broeck_thermodynamic_2005}. When time-reversal symmetry is broken, the second law of thermodynamics only imposes the weaker bound $\eta_{\rm{maxP},l}\leq\eta_c$~\cite{benenti_thermodynamic_2011}. However, current conservation in multi-terminal setups gives rise to additional constraints that lead to $\eta_{\rm{maxP},l}<\eta_c$ with the precise bound depending on the number of terminals~\cite{brandner_multi-terminal_2013}. We remark that for a three-terminal setup as considered here one has $\eta_{\rm{maxP},l}\leq\eta_c/2$.
%The possibility to tune the parameters $x_{ij}$ has been suggested to improve the efficiency at maximum power~\cite{benenti_thermodynamic_2011,brandner_multi-terminal_2013} over the bound of $\eta_c/2$ established for time-reversal symmetric systems~\cite{van_den_broeck_thermodynamic_2005}. For an large number of terminals, these bounds can be close to the Carnot efficiency $\eta_c$. The described tunability of the asymmetry $x_{ij}$ in quantum Hall systems postulates our system for tests the bounds of the thermoelectric efficiency. We explore this posibility in the next sections.}

\subsection{Heat rectification}
The effect of chirality is not only restricted to the propagation of charge. It also affects the heat currents. This effect is parametrised by the heat rectification coefficient
\be
{\cal R}_{ij}=\frac{{\cal L}_{ij}^{hT}}{{\cal L}_{ji}^{hT}}
\ee
that relates the thermal responses to opposite temperature gradients. ${\cal R}_{ij}=1$ implies no heat rectification. A system with ${|}{\ln}{\cal R}_{ij}{|}\gg1$ works as an efficient heat diode. For the longitudinal term, we get the simple expression
\be
{\cal R}_{12}=\left[1-\frac{e\lambda^{(2)}}{h\kBT^2}\left(\frac{1}{G_1S_1}+\frac{1}{G_2S_2}\right)j^{(2)}+\frac{e^2\kBT}{hG}j^{(3)}\right]^{-1}.
\ee
As in the chiral terms ${\cal X}_l$, rectification relies on the coherent propagation between terminals 1 and 2 through the integrals $j^{(n)}$.  We emphasize that this is a three-terminal effect due to chirality. In the absence of the probe terminal or of the magnetic field, we recover ${\cal R}_{12}=1$. 

Rectification can also occur involving fluxes from the probe terminal. In this case, we obtain the simple expressions:
\bea
\label{r13}
{\cal R}_{13}&=\left[1-\frac{j^{(2)}m^{(2)}-j^{(3)}m^{(1)}}{g_1^{(2)}m^{(2)}-g_1^{(3)}m^{(1)}}\right]^{-1},\\
\label{r23}
{\cal R}_{23}&=1-\frac{j^{(2)}m^{(2)}-j^{(3)}m^{(1)}}{g_2^{(2)}m^{(2)}-g_2^{(3)}m^{(1)}},
\eea
where $m^{(n)}=g_1^{(n)}+g_2^{(n)}-j^{(n)}$. Here again, there is no rectification if $j^{(n)}=0$.

In the following, we illustrate the formal results derived in this section with two paradigmatic configurations with experimental relevance: We consider the cases where the junctions 1 and 2 consist of quantum point contacts or tunneling resonances, with the coupling of the probe terminal being transparent. As it will become clear, the response of the system will be completely different in each case. Quantum point contacts allow for totally switching on and off the transport channels by an appropriate gating of the system. Resonances behave as energy filters: transport is closed for all energies except those close to the resonances. This is known to increase the thermoelectric efficiency but limits the thermal rectification properties.  
%discuss the response of the configuration with a transparent probe for different scattering configurations of junctions 1 and 2. We will consider the cases where they consist of quantum point contacts or resonances.  

%%%%%%%%%%%%%%%%%%%%%
%\section{Particular configurations}
%\label{particular}
%%%%%%%%%%%%%%%%%%%%%

%%%%%%%%%%%%%%%%%%%%%
\section{Quantum point contacts}
\label{sec:qpc}
%%%%%%%%%%%%%%%%%%%%%
Let us consider a typical configuration where the scattering at the conducting terminals is introduced by quantum point contacts.
In the case where the potential defining the quantum point contact can be approximated by a saddle point potential, the transmission coefficient is given by a broadened step function~\cite{fertig_transmission_1987,buttiker_quantized_1990}
\be
{\cal T}_{\text{QPC},l}(E)=[1+e^{-2\pi(E-E_l)/\hbar\omega_{0,l}}]^{-1},
\ee
whose position and width are given by $E_l$ and $\omega_{0,l}$, respectively. They can be easily controled by means of gate voltages. Importantly for our purposes here, they can be switched from being open or closed by tuning the step energy $E_l$ far below or above the Fermi energy, respectively.

\begin{figure}[t]
\begin{center}
\includegraphics[width=0.8\linewidth,clip]{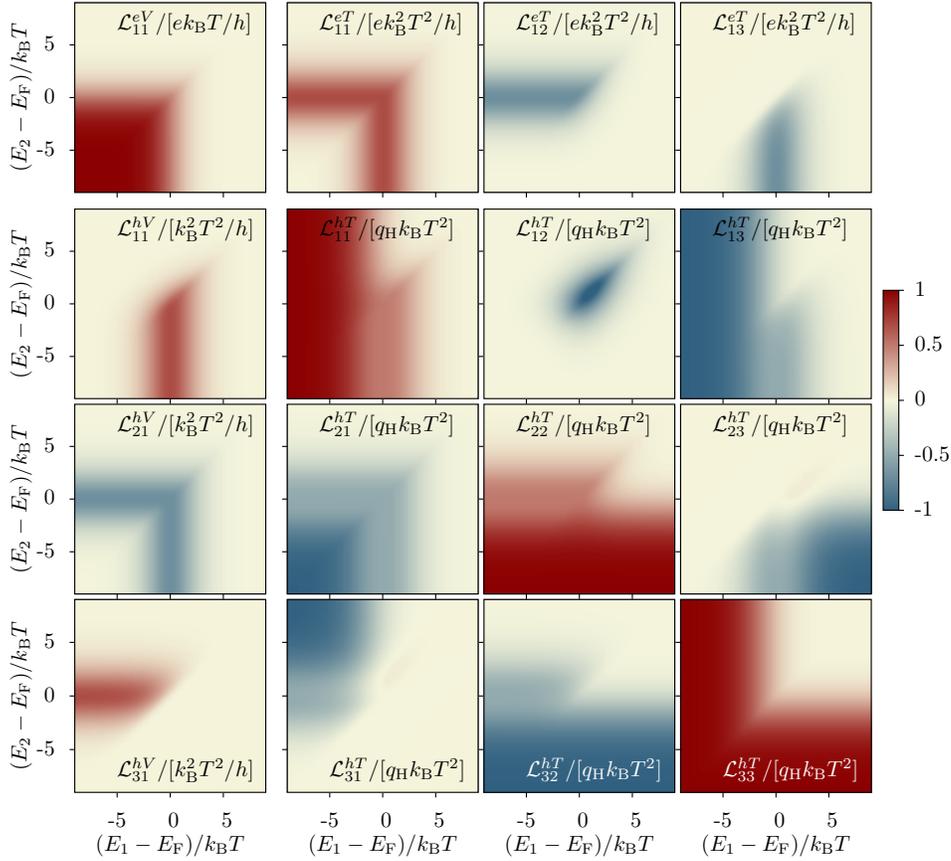}
\end{center}
\caption{\label{onsager} Representation of the Onsager matrix ${\mathbf{\cal L}}$. We consider two constrictions in front of the conducting terminals in the form of QPCs with threshold energies $E_1$ and $E_2$, and $\hbar\omega_0=\kBT$. The reciprocity relations between off-diagonal elements are recovered upon reversing the magnetic field. For this configuration, the Seebeck response to a temperature bias in terminals 2 and 3 are sensitive to chiral motion as they depend on which of the constrictions is present. However, only ${\cal L}_{13}^{eT}$ shows a chiral behaviour irrespective of the sign of the magnetic field, as ${\cal L}_{13}^{eT}(-B)=e{\cal L}_{13}^{hV}(B)$.}
\end{figure}

In \fref{onsager}, we plot the different terms of the Onsager matrix given in Eqs.~\eref{L11et} to \eref{L31hv} by tuning the step energy of the contacts in terminals 1 and 2. For the charge conductance, we get the expected single-channel quantization~\cite{van_wees_quantized_1988} as the two constrictions are opened. The coefficients related to thermal conductances also present step-like dependences relying on the opening/closing of the junctions. The step height is given by the quantum of heat conductance $q_{\rm H}=\pi^2\kB^2T/(3h)$~\cite{sivan_multichannel_1986}. The longitudinal coefficients ${\cal L}_{11}^{hT}$ and ${\cal L}_{22}^{hT}$ present a double-step structure due to the competition of $\hbar\omega_0$ and $\kBT$. 

\begin{figure}[t]
\begin{center}
\includegraphics[width=0.8\linewidth,clip]{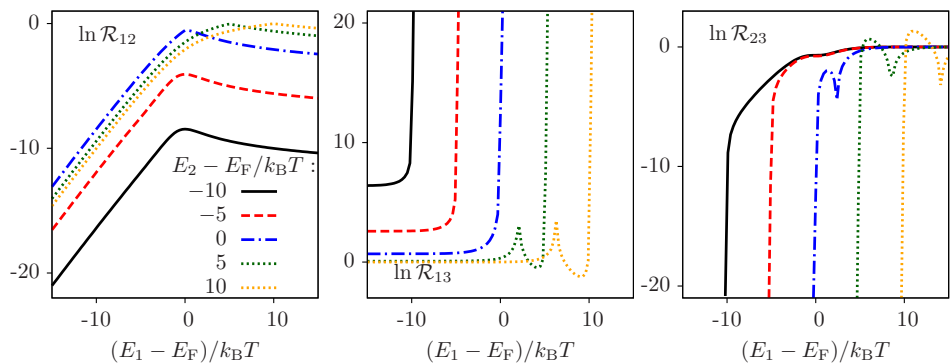}
\end{center}
\caption{\label{rij} Heat rectification coefficient $\ln{\cal R}_{ij}$ corresponding to the configuration of \fref{onsager}. As the energy $E_1$ is becomes positive and larger than $\kBT$, the contact 1 gets closed. The different curves correspond to different positions of the contact in terminal 2. The longitudinal coefficient ${\cal R}_{12}$ is exponentially suppressed when junction 1 is open. The crossed coefficients ${\cal R}_{13}$ and ${\cal R}_{23}$ show divergences when one of the corresponding coefficients changes sign. }
\end{figure}
Remarkably the presence of the probe terminal causes a pronounced heat rectification. This is clear by looking at the asymmetry between the longitudinal terms ${\cal L}_{12}^{hT}$ and ${\cal L}_{21}^{hT}$. The most remarkable case is when the two junctions are open. Then, heat flows longitudinally along the lower edge if terminal 1 is hot, but is suppressed if the temperature bias is applied to terminal 2. In the latter case, due to chiral propagation from terminal 2 to 3, the non-equilibrium carriers injected from terminal 2 are {\it all} absorbed by the probe, where they thermalise due to inelastic scattering. Thus, the heat injected in terminal 2 flows into the probe. Therefore, ${\cal R}_{12}$ becomes exponentially suppressed  when contact 1 is opened, cf. \fref{rij}. 

The crossed rectification coefficients present a strong deviation from 1 when the system is close to the symmetric configuration with $E_1=E_2$ even if both contacts are still open. This is because one of the junctions preselects the scattering at the other one. Let us consider first the case of ${\cal R}_{13}$. When $E_1>E_2$, the non-equilibrium carriers that are not reflected in junction 1 will most probably neither be reflected at junction 2. Therefore heat flow into the probe is strongly suppressed and ${\cal L}_{31}^{hT}=0$. Similarly for the case of ${\cal R}_{23}$, when $E_1<E_2$ electrons injected from the probe will either be transmitted at junction 1 or reflected at both junctions. Then, none of them reach terminal 2 and ${\cal L}_{23}^{hT}=0$. 

We also observe a peak structure appearing for some of the curves. They occur because ${\cal L}_{13}^{hT}$ and ${\cal L}_{32}^{hT}$ can eventually change sign. This is the case for quantum point contacts with $\hbar\omega_0$ not much larger than $\kBT$. As a clear example, let us consider the limit where $\hbar\omega_0\ll\kBT$, so the scattering can be approximated by step functions. In that case, when $E_2<E_1$, we have $j^{(n)}=g_1^{(n)}$, so ${\cal R}_{13}$ diverges and ${\cal R}_{23}=0$, as it follows from Eqs.~\eref{r13} and \eref{r23}. Our system therefore works as a versatile and extremely efficient heat diode.

We now turn to the thermoelectric coefficients. They are finite when at least one of the junctions is noisy, cf. \fref{onsager}. This is expected from the thermopower of quantum point contacts~\cite{streda_quantised_1989,molenkamp_quantum_1990,molenkamp_peltier_1992}. The presence of a finite thermoelectric response depends on where the charge current is measured and to which terminal the temperature bias is applied. For instance, ${\cal L}_{11}^{eT}$ is finite when either junction is constricted. This is not the case if terminal 2 is hot, where no current flows if the junction in front of it is open (for $E_2-E_\text{F}\ll\kBT$). For the Peltier coefficients ${\cal L}_{ij}^{hV}$, the situation is the opposite. As discussed in section~\ref{scattering}, this is because the chiral terms contribute to only one of the longitudinal Seebeck or Peltier coefficients for each sign of the magnetic field, cf. Eqs.~\eref{L11et} to \eref{L31hv}. 

The crossed thermoelectric responses present a more robust dependence on chirality. Either the Seebeck or the Peltier coefficient will be suppressed when one of the junctions is open for both signs of the magnetic field~\cite{sanchez_chiral_2014}. Thus, the asymmetry factors $x_{ij}$ can be tuned from zero to $\pm\infty$ by acting on the contact gates. 

\begin{figure}[t]
\begin{center}
\includegraphics[width=0.7\linewidth,clip]{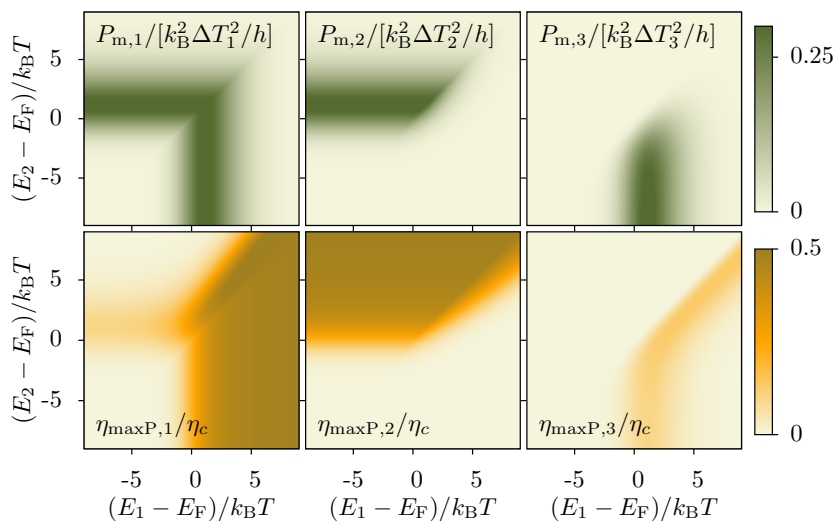}
\end{center}
\caption{\label{pmqpc} Maximum power $P_{{\rm m},l}$ and efficiency at maximum power $\eta_{{\rm maxP},l}$, when a temperature bias $\Delta T_l$ is applied to terminal $l$. We consider two constrictions in front of the conducting terminals in the form of QPCs with threshold energies $E_1$ and $E_2$, and $\hbar\omega_0=10^{-2}\kBT$.}
\end{figure}

In \fref{pmqpc} we show the thermoelectric performance of a heat engine based on these effects. We compare the maximum power $P_{{\rm m},l}$ and the efficiency at maximum power $\eta_{{\rm maxP},l}$ for different cases depending on to which terminal, $l$ the heat source is coupled. For quantum point contacts, the optimal configuration is that with $\hbar\omega_0\ll\kBT$. For the longitudinal cases, the maximal $\eta_{{\rm maxP},l}\approx\eta_c/2$ is reached when the junctions are close to pinch-off. Unfortunately, the extracted power is then suppressed. One has therefore to find a compromise between a finite power extraction and a high enough efficiency. For instance, configurations with the heat source in terminals 1 and 3 have the peculiarity that some maxima of $P_{{\rm m},l}$ coincide with a maximum of $\eta_{{\rm maxP},l}$.

%%%%%%%%%%%%%%%%%%%%%
\section{Resonant tunneling}
\label{sec:restunn}
%%%%%%%%%%%%%%%%%%%%%
Resonant contacts are useful for thermoelectrics because they are good energy filters~\cite{edwards_cryogenic_1995,prance_electronic_2009,humphrey_reversible_2002,jordan_powerful_2013,sothmann_powerful_2013}. In a quantum Hall bar, they can be formed by a series of point contacts (as sketched in the right arm of \fref{scheme}), an antidot, or additional probes~\cite{pascher_resonant_2014}. The scattering at these junctions can be approximated by Breit-Wigner resonances~\cite{buttiker_coherent_1988}. The transmission probabilities are Lorentzian distributions defined by their resonant energy $E_l$ and their width $\Gamma_l$
\be
{\cal T}_{\text{res},l}(E)=\frac{\Gamma_l^2}{4(E-E_l)^2+\Gamma_l^2},
\ee
where we have assumed that the coupling of the resonance to the outer channels is symmetric.

\begin{figure}[t]
\begin{center}
\includegraphics[width=0.8\linewidth,clip]{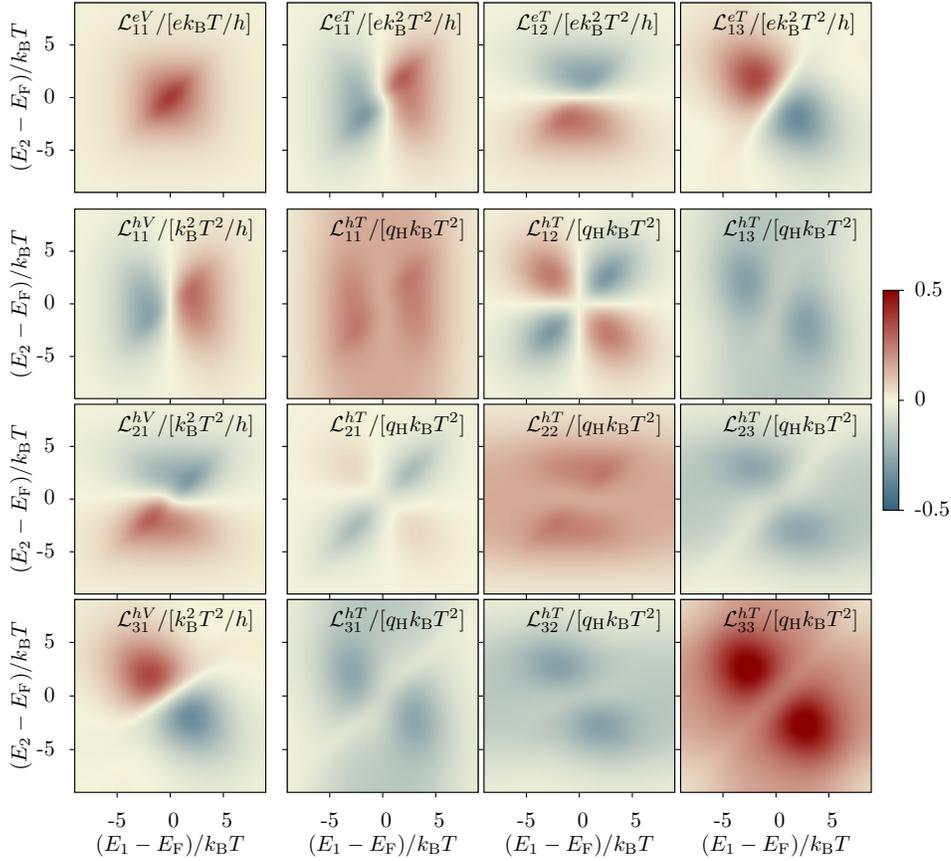}
\end{center}
\caption{\label{onsagerres} Representation of the Onsager matrix ${\mathbf{\cal L}}$. We consider two constrictions in front of the conducting terminals in the form of resonances with threshold energies $E_1$ and $E_2$, and $\Gamma_1=\Gamma_2=2\kBT$. }
\end{figure}
In \fref{onsagerres} we show a representation of the Onsager matrix as the energy of the resonances $E_1$ and $E_2$ are tuned around the Fermi energy $E_{\rm F}$. The charge conductance is peaked when the two resonances align with the Fermi level. All the other coefficients present a more complicated lobe structure.

It can readily be seen that the off-diagonal thermal coefficients show a similar dependence as their transpose element. Nevertheless, large deviations of the rectification coefficient from unity appear which are restricted to particular configurations: For the longitudinal term, ${\cal R}_{12}$, this is the case when one of the resonances crosses the Fermi level, such that the corresponding thermopower $S_l$ vanishes (not shown). Around this value, ${\cal R}_{12}$ also changes sign. Thus, even if heat rectification is suppressed by the presence of scatterers, an efficient and tuneable heat diode can be built based on energy filters. However, the amount of rectified heat is limited by the width of the resonances. Such large deviations are not present in the crossed rectification coefficients ${\cal R}_{13}$ and ${\cal R}_{23}$, which are maximal at symmetric configurations with $E_1\approx E_2$.
% unless the width of one of the resonances is so large that its junction can be considered open.

On the other hand, energy filtering is essential for high thermoelectric efficiencies~\cite{mahan_best_1996} as it enables a tight charge-energy coupling. The coefficient ${\cal L}_{13}^{eT}$ responsible for energy harvesting from the hot probe is maximal when the two levels are symmetrically placed around the Fermi energy. In that configuration, each contact contributes either to the transport of electrons or of holes. The electron-hole excitations created in the heat source are split at the different junctions and current is generated between the conductor terminals. This mechanism has been proposed for powerful and efficient energy harvesting with systems of quantum dots~\cite{jordan_powerful_2013} and quantum wells~\cite{sothmann_powerful_2013}. Differently in the present case, the crossed response is in general finite in the symmetric configurations with $E_1=E_2$ due to the chirality ${\cal X}_l$, cf. \fref{onsagerres}.

The longitudinal responses ${\cal L}_{1l}^{eT}$ vanish close to the condition where the resonance in front of the hot terminal becomes particle-hole symmetric, with $E_l=E_{\rm F}$. The sawtooth like oscillation of the thermovoltage is typical of resonant systems~\cite{beenakker_theory_1992,staring_coulomb-blockade_1993,dzurak_thermoelectric_1997,svensson_lineshape_2012}: the thermopower changes sign as the resonance crosses the Fermi level for filtering either the electron-like or the hole-like carriers. For the Peltier coefficients, ${\cal L}_{l1}^{hV}$ the same is true for the terminal at which the heat current is measured. These different conditions introduce clear divergences in the asymmetry factors $x_{12}$ and $x_{13}$.

Let us now discuss the thermoelectric performance. In the limit of narrow resonances with $\Gamma_l\ll\kBT$ we can get simple analytical results as the transmission probabilities are approximated by ${\cal T}_l(E)\rightarrow\Gamma_l\delta(E-E_l)$. Then, from Eqs.~\eref{L33ht} and \eref{L13et} it can be easily shown that the crossed thermoelectric terms satisfy
\be
{\cal L}_{13}^{eT}=\frac{e}{E_1-E_2}{\cal L}_{33}^{hT},
\ee
i.e. the generated current and the injected heat are proportional to each other. Their ratio is given only by the elementary charge $e$ and the level detuning $E_1-E_2$ defined by the resonances. This is the tight charge-energy coupling in which thermoelectric Carnot efficiency is attained. Indeed, we obtain that $\eta(V_s)=\eta_c$ for the stall potential $V_s=(E_1-E_2)/e$ and $\eta_{{\rm maxP},3}=\eta_c/2$. These results recover what is found for systems of quantum dots~\cite{jordan_powerful_2013}. Note that in this limit, the integrals $j^{(n)}=0$ (except for the particular condition when $E_1=E_2$) such that the chiral contributions ${\cal X}_l$ are not present. Unfortunately, the extracted power decreases with the resonance width, so it is vanishingly small in this limit.

\begin{figure}[t]
\begin{center}
\includegraphics[width=0.7\linewidth,clip]{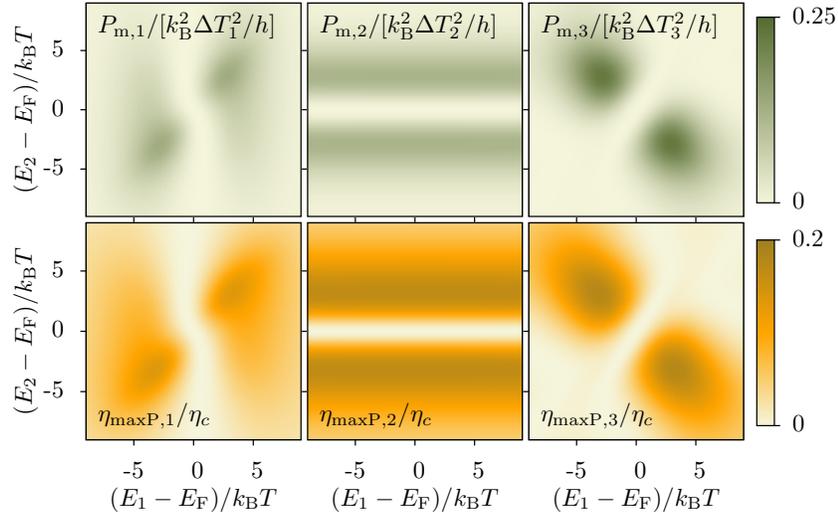}
\end{center}
\caption{\label{pmres} Maximum power $P_{{\rm m},l}$ and efficiency at maximum power $\eta_{{\rm maxP},l}$, when a temperature bias $\Delta T_l$ is applied to terminal $l$. We consider two constrictions in front of the conducting terminals in the form of resonances with threshold energies $E_1$ and $E_2$, and optimized widths in order to give the maximum $P_{\rm m}$: $\Gamma_1=\Gamma_2=3.32\kBT$ (for $l=1$), $\Gamma_1\gg\Gamma_2=2.17\kBT$ (for $l=2$), and $\Gamma_1=\Gamma_2=1.975\kBT$ (for $l=3$).}
\end{figure}

In order to extract larger currents, the parameters of the system need to be optimized. In \fref{pmres}, we plot the maximum power and the corresponding efficiency for $\Gamma_1$ and $\Gamma_2$ chosen such that $P_{{\rm m},l}$ is maximized. We compare the performances obtained for the different terminals being hot. All of  them show a coincidence between the regions where $P_{{\rm m},l}$ and $\eta_{{\rm maxP},l}$ are large. The crossed response ($l=3$) presents the best performance as a larger power is generated at a larger efficiency. Note the very different conditions obtained for each configuration: while for a hot terminal 1, it is preferred that $E_1=E_2$, the crossed response is enhanced when $E_1=-E_2$. Differently, a hot terminal 2 has a most favourable configuration when $\Gamma_1\gg\kBT$ such that this junction is effectively open.

%%%%%%%%%%%%%%%%%%%%%
\section{Conclusions}
\label{sec:conclusions}
%%%%%%%%%%%%%%%%%%%%%
We have investigated in detail the charge and heat current responses of a three-terminal quantum Hall conductor. The chiral propagation of electrons along the edge states manifests itself in the thermal and thermoelectric Onsager coefficients. This contribution introduces deviations from configurations with no magnetic field which are responsible e.g. for a divergence of the crossed Seebeck to Peltier ratio and the possibility to harvest energy from symmetric configurations~\cite{sanchez_chiral_2014}. The longitudinal transport coefficients correspond to a two-terminal configuration with a voltage probe added to it. We showed that the presence of the probe has tremendous consequences on the thermal rectification coefficient which allow us to propose our system as both an efficient heat engine and an efficient heat diode.
% We are confident in the relevance of quantum Hall heat engines for practical applications even at room temperature~\cite{novoselov_room_2007}. 

We have explored two different configurations consisting of quantum point contacts and resonant junctions. A system based on resonances presents peaks in the heat rectification which can be tuned by gate voltages. In this case, the effect of chirality of the carriers is suppressed for the configuration relevant for heat conversion purposes. Therefore, the thermoelectric performance of resonant junctions is similar to that predicted in the absence of a magnetic field. Very differently, quantum point contacts allow for opening and closing the transport channels. As a consequence, the system behaves as an ideal and versatile thermal diode and an efficient and powerful heat engine. 

\ack
We acknowledge financial support from the Spanish MICINN Juan de la Cierva program and MAT2011-24331, the COST Action MP1209 and the Swiss National Science Foundation. 

%%%%%%%%%%%%%%%%%%%%%
\appendix
%%%%%%%%%%%%%%%%%%%%%
\section{Closed loop configuration. A dot}
\label{sec:closed}
%%%%%%%%%%%%%%%%%%%%%
\begin{figure}[t]
\begin{center}
\includegraphics[width=0.65\linewidth,clip]{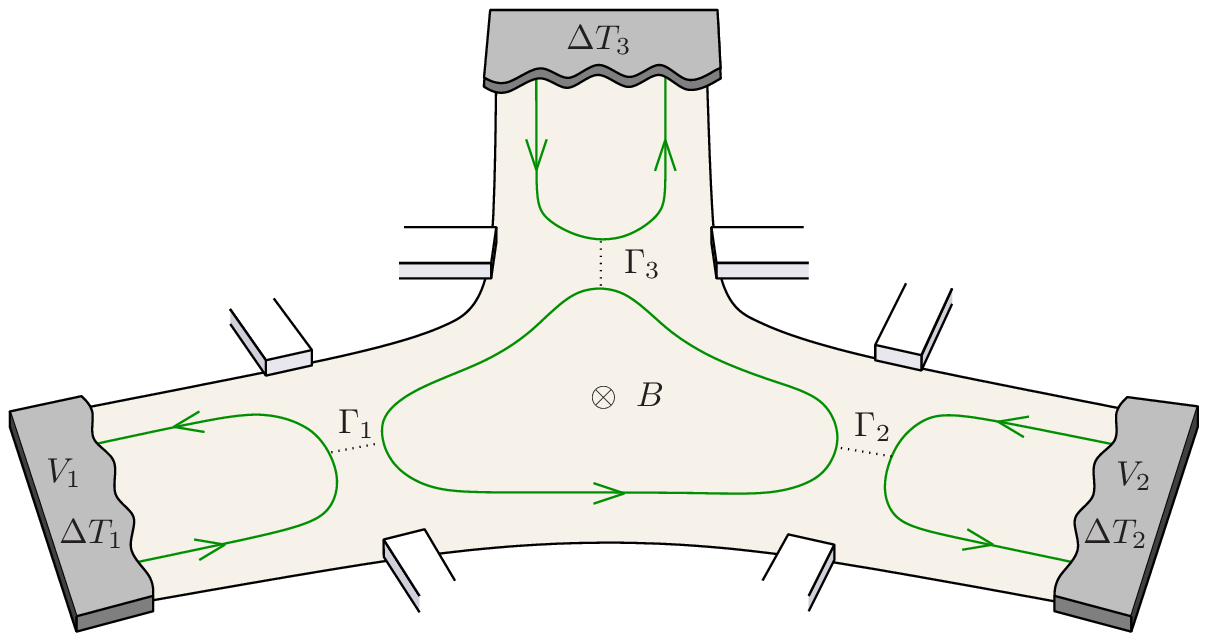}
\end{center}
\caption{\label{loop} Three-terminal quantum Hall bar with reflexion at the three junctions. A closed loop is then formed in the center of the structure which gives rise to resonant tunneling connecting the three terminals. The couplings are given by $\Gamma_i$.}
\end{figure}

We get further insight on the importance of chiral propagation in the case where the three junctions are noisy. This system can be mapped to an edge state forming a closed loop which is coupled to the three terminals, as sketched in \fref{loop}. Then, the interference of multiple scattering trajectories within the loop gives rise to a Fabry-Perot like resonant behaviour. The transmission coefficients can be described by a Breit-Wigner resonance~\cite{buttiker_negative_1988}:
\bea
{\cal T}_{i\leftarrow i}&=1-\frac{\Gamma_i(\Gamma-\Gamma_i)}{\Delta}\\
{\cal T}_{i\leftarrow j}&=\frac{\Gamma_i\Gamma_{j-1}}{\Delta},
\eea
with $\Delta=(E-E_r)^2+\Gamma^2/4$, $E_r$ being the resonance energy and $\Gamma=\sum_i\Gamma_i$, its width. Controlled resonances can be realized in experiment~\cite{pascher_resonant_2014}. 

Considering again the case that terminal 3 is a voltage probe, the electric response is given by ${\cal L}_{11}^{eV}=\kBT G_{\rm d}$, with
\be
G_{\rm d}=g_{12}+\frac{g_{13}g_{23}}{g_{13}+g_{23}}.
\ee 
Here, we have defined the partial conductances $g_{ij}=e^2(\kBT h)^{-1}\int dE\Gamma_i\Gamma_j\Delta^{-1}\xi(E)$. By defining the partial thermovoltages $s_{ij}=e^2(\kB T^2 hg_{ij})^{-1}\int dEE\Gamma_i\Gamma_j\Delta^{-1}\xi(E)$, we get the crossed thermoelectric response 
\be
\label{L13etdot}
{\cal L}_{13}^{eT}=\kB T^2(G_{\rm d}-g_{12})(s_{23}-s_{13})
\ee
when terminal 3 is hot. It is interesting to compare Eqs.~\eref{L13et} and \eref{L13etdot}. They both include a term depending on the difference of the thermovoltage related to the junctions in the conductor terminals. In \eref{L13etdot}, the importance of how each terminal is connected to the heat source is emphasized. Notably in this case, the chiral term ${\cal X}_l$ is not present. The reason is that every terminal is connected to each other via the central loop and this contribution is cancelled. 

By inspection of Eq.~\eref{L13etdot}, it is clear that at least one of the partial widths $\Gamma_1$ or $\Gamma_2$ must be energy dependent in order to have a finite response. In this case, the particle-hole asymmetry introduced by the resonance is not sufficient. No requirement is put on $\Gamma_3$, i.e. the response does not rely on the way that the probe injects heat into the system.

%%%%%%%%%%%%%%%%%%%%%
\section{Closed loop configuration. An antidot}
\label{sec:antidot}
%%%%%%%%%%%%%%%%%%%%%
\begin{figure}[t]
\begin{center}
\includegraphics[width=0.65\linewidth,clip]{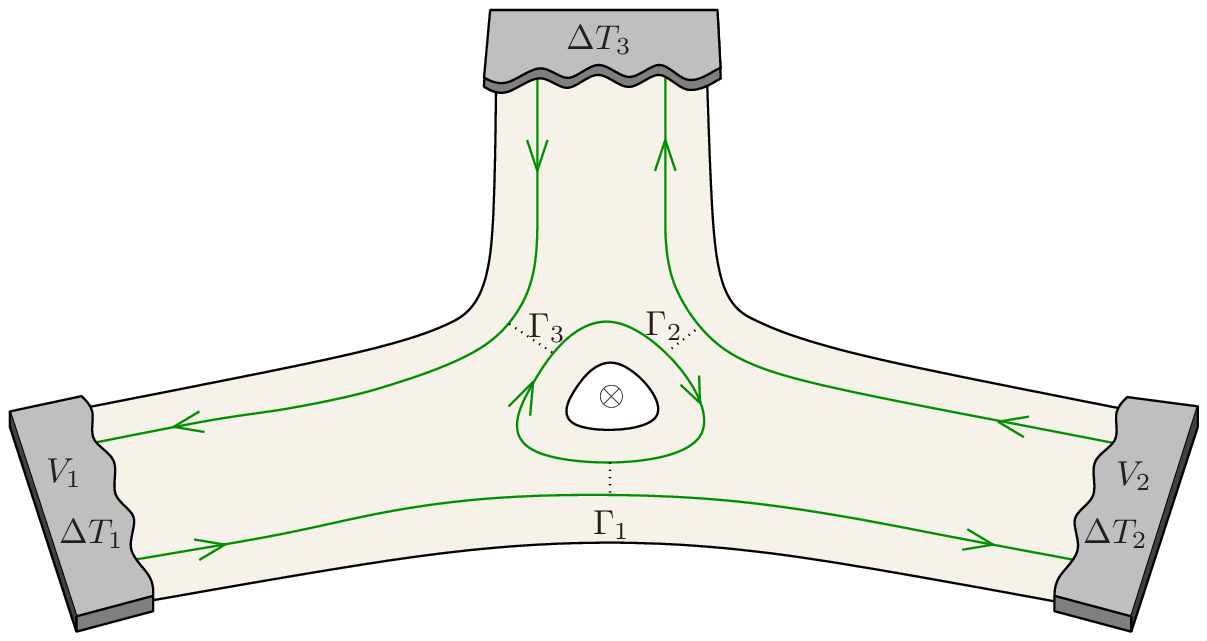}
\end{center}
\caption{\label{antidot} Three-terminal quantum Hall bar with an antidot. Electrons describe clockwise orbits around the antidot, whereas the propagation in the rest of the sample is anticlockwise . The couplings are given by $\Gamma_i$.}
\end{figure}

A different effect is given by the presence of an antidot. As shown in \fref{antidot}, a closed loop is also formed around the antidot. Differently to the previous case presented in \ref{sec:closed}, the chirality of propagation around the loop is opposite to the rest of the sample. The transmission coefficients are in this case given by:
\bea
{\cal T}_{i\leftarrow j}&=\frac{\Gamma_i\Gamma_{j-1}}{\Delta}\quad{\rm (if\  }j\ne i-1{\rm )}\\
{\cal T}_{i+1\leftarrow i}&=1-\frac{\Gamma_i(\Gamma-\Gamma_i)}{\Delta}\quad {\rm (otherwise)},
\eea
where all indices are to be taken modulo 3~\cite{buttiker_negative_1988}.

In this case, the crossed Seebeck coefficient can be written as:
\be
{\cal L}_{13}^{eT}=\kB T^2\frac{g_{13}g_{23}}{g_0-g_{23}}(s_{23}-s_{13})+\kB T^2\frac{g_{0}g_{13}}{g_0-g_{23}}s_{13},
\ee
in terms of a difference of left and right thermovoltages and an additional term. In analogy to the chiral term in Eq.~\eref{L13et}, this term depends now only on the coupling between the hot source and terminal 1. A chiral-dependent behaviour is recovered because the electrons injected from terminal 3 which are reflected at constriction 3 are unavoidably absorbed by terminal 1. 

Different from the closed-loop case with a dot considered above, the energy filtering of the antidot is enough to generate a thermoelectric response, even if the partial widths $\Gamma_i$ are all constant or equal. 

\section*{References}

% \bibliographystyle{iopart-num}
% \bibliography{/home/bjoern/LaTeX/Bibtex/Meine_Bibliothek}

\providecommand{\newblock}{}

\end{document}